\documentclass[conference,10pt]{IEEEtran}

\usepackage{cite}
\usepackage{overpic}
\usepackage{url}
\usepackage[dvipsnames,svgnames]{xcolor}
\usepackage{graphicx}
\usepackage{psfrag}
\DeclareGraphicsExtensions{.xps}
\usepackage{epstopdf} 
\usepackage[cmex10]{amsmath}
\usepackage{amssymb}
\usepackage{amsthm}

\def\tr{\mathrm{tr}}

\newtheorem{theorem}{Theorem}

\newtheorem{remark}{Remark}
\newtheorem{assumption}{Assumption}

\usepackage[textwidth=30mm]{todonotes}

\newcommand{\vect}[1]{\mathbf{#1}}

\def\tr{\mathrm{tr}}

\def\Htran{\mbox{\tiny $\mathrm{H}$}}

\def\CN{\mathcal{N}_{\mathbb{C}}} 

\begin{document}

\IEEEoverridecommandlockouts

\title{Large-System Analysis of Massive MIMO with Optimal M-MMSE Processing\vspace{-0.5cm}}

\author{
\IEEEauthorblockN{Luca Sanguinetti\IEEEauthorrefmark{1}, Emil Bj{\"o}rnson\IEEEauthorrefmark{2}, Abla Kammoun\IEEEauthorrefmark{3}}
\IEEEauthorblockA{\IEEEauthorrefmark{1}\small{Dipartimento di Ingegneria dell'Informazione, University of Pisa, Pisa, Italy}}
\IEEEauthorblockA{\IEEEauthorrefmark{2}\small{Department of Electrical Engineering (ISY), Link\"{o}ping University, Link\"{o}ping, Sweden}}
\IEEEauthorblockA{\IEEEauthorrefmark{3}\small{Electrical Engineering Department, King Abdullah University of Science and Technology, Thuwal, Saudi Arabia}}
\vspace{-1cm}}
\maketitle

\begin{abstract}
We consider the uplink of a Massive MIMO network with $L$ cells, each comprising a BS with $M$ antennas and $K$ single-antenna user equipments. Recently, \cite{BHS18A} studied the asymptotic spectral efficiency of such networks with optimal multicell minimum mean-squared error (M-MMSE) processing when $M\to \infty$ and $K$ is kept fixed. Remarkably, \cite{BHS18A} proved that, for practical channels with spatial correlation, the spectral efficiency grows unboundedly, even with pilot contamination. In this paper, we extend the analysis from \cite{BHS18A} to the alternative regime in which $M,K\to \infty$ with a given ratio. Tools from random matrix theory are used to compute low-complexity approximations which are proved to be asymptotically tight, but accurate for realistic system dimensions, as shown by simulations.\end{abstract}

\section{Introduction}
Massive MIMO is a wireless network technology where the base stations (BSs) are equipped with a very large number $M$ of low-power, fully digitally controlled, and physically small antennas to serve a multitude of user equipments (UEs) by spatial multiplexing \cite{marzetta2010noncooperative}. 
A rigorous and mature theory for Massive MIMO has been developed in recent years, as underlined by the recent textbooks \cite{Marzetta2016a} and \cite{massivemimobook}.

In industry, exciting developments occurred in 2018. The technology has been integrated into the 5G New Radio standard \cite{Parkvall2017a}, and the first 64-antenna 
Massive MIMO BSs have been added to the Ericsson  AIR, Huawei  AAU, and Nokia  AirScale product lines and commercially deployed \cite{Sprint2018feb}. 
 This manifests that Massive MIMO is no longer a promising concept but a reality for cellular networks (below 6 GHz). 

In academia, Massive MIMO was originally characterized by the ``Marzetta limit'' where $M\to\infty$ while the number $K$ of UEs is fixed \cite{marzetta2010noncooperative}. 
This limit is different from the traditional ``large-system limit'' where $M,K\to\infty$ with a fixed ratio. The Marzetta limit has the practical benefit that the $K$ pilot resources required for channel estimation remain finite even in the asymptotic limit. The Massive MIMO capacity was first believed to be upper limited by the coherent interference created by \emph{pilot contamination} (i.e., reuse of pilots across cells). However, this issue was recently resolved in \cite{BHS18A,Neumann2018,SanguinettiGC2018}. More precisely, \cite{BHS18A} proved that, with optimal multicell minimum mean-squared error (M-MMSE) processing, the capacity grows unboundedly as $M\to\infty$. The only requirement is that the channel correlation matrices of the contaminating users are asymptotically linearly independent. This was not the case in Marzetta's original paper \cite{marzetta2010noncooperative}, but channel measurements show that it is likely the case in practice  \cite{Gao2015b}.
 Similar results can be obtained by using a generalized matched filter \cite{Neumann2018,SanguinettiGC2018}. 


Any practical system will operate with a finite $M$ and $K$. Therefore, the purpose of asymptotic analysis is not the limit itself but to understand the capacity scaling behavior and obtain tight low-complexity performance approximations. To this end, we should choose between the Marzetta limit and traditional large-system limit depending on whether $M/K$ will be nearly infinite or small in practice. Since the sum capacity is often maximized when $M/K$ is fairly small \cite{massivemimobook,Bjornson2016a}, the traditional large-system limit is still of interest.

In this paper, we extend the asymptotic analysis from \cite{BHS18A,Neumann2018}, which considers the Marzetta limit, to the traditional regime in which $M,K\to\infty$ with $\liminf M/K>0$. 
 To the best of our knowledge, only suboptimal schemes such as maximum ratio, zero-forcing, and single-cell MMSE processing are considered in prior work; see e.g., \cite{Hoydis2013,Sadeghi2017}. M-MMSE is investigated in \cite{EmilEURASIP17} but only for uncorrelated Rayleigh fading channels. 
This paper fills the gap by providing an analytical framework that allows evaluating the performance of a Massive MIMO network with M-MMSE for practically large numbers of $M$ and $K$, without the need of carrying out computationally demanding Monte Carlo simulations. Moreover, it provides novel insights into the achievable performance when using M-MMSE processing.

\begin{figure*}
 \setcounter{equation}{12}
\begin{align} \gamma_{jk}&= \hspace{-1cm}\underbrace{\hat{\vect{h}}_{jjk}^{\Htran}{\bf A}_{jk}^{-1}\hat{\vect{h}}_{jjk}}_{\text{Independent of pilot contaminating channels}} \hspace{-1.cm} - \hspace{0.2cm}
  \underbrace{\hat{\vect{h}}_{jjk}^{\Htran}  {\bf A}_{jk}^{-1}\hat{\vect{H}}_{jk}^{[j]}\left(  {\bf I}_{L-1}+ {\big(\hat{\vect{H}}_{jk}^{[j]}\big)}^{\Htran}{\bf A}_{jk}^{-1}\hat{\vect{H}}_{jk}^{[j]}\right)^{-1}\!\!\!{\big(\hat{\vect{H}}_{jk}^{[j]}\big)}^{\Htran}{\bf A}_{jk}^{-1}\hat{\vect{h}}_{jjk}}_{\text{Loss due to the correlation among pilot contaminating channels}}\label{eq:appendixc_7}
\end{align}\vspace{-0.2cm}
\hrule\vspace{-0.3cm}
\end{figure*}
 \setcounter{equation}{0}
 
\section{Massive MIMO System Model}
\label{section:multi-user}

We consider a Massive MIMO network with $L$ cells, each comprising a BS with $M$ antennas and $K$ single-antenna UEs. We consider a block-fading system
model where each channel takes one realization in a coherence
block of $\tau_c$ channel uses and independent realizations across
blocks. There are $K$ mutually orthogonal pilots and the $k$th UE in each cell uses the same pilot. 
Following the notation from \cite{Hoydis2013}, the received signal ${\bf y}_j \in \mathbb{C}^{M}$ at BS $j$ is
\begin{equation}
{\bf y}_j = \sum_{l=1}^{L}  \sum_{i=1}^{K} \sqrt{\rho} \vect{h}_{jli} x_{li} + \vect{n}_j
\end{equation}
where $\rho$ is the normalized transmit power, $x_{li}$ is the signal from UE $i$ in cell $l$, $\vect{n}_j \sim \CN (\vect{0}, \vect{I}_{M})$ is the normalized independent receiver noise at BS~$j$, and $\vect{h}_{jli} \sim \CN (\vect{0}, \vect{R}_{jli})$ is the block-fading channel from this UE to BS $j$. The covariance/correlation matrix $\vect{R}_{jli} \in \mathbb{C}^{M \times M}$ accounts for the large-scale fading, including pathloss and spatial correlation \cite{massivemimobook}. These matrices are assumed to be known, but practical estimation methods are found in \cite{BSD16A,NeumannJU17,UpadhyaJU17}.


\subsection{Channel Estimation and Spectral Efficiency}
Using a total uplink pilot power of $\rho^{\rm{tr}}$ per UE and standard MMSE estimation techniques \cite{Hoydis2013}, BS $j$ obtains the estimate of $\vect{h}_{jli}$ as
\begin{align}\label{channel_estimates}
\!\!\!\hat{\vect{h}}_{jli} = \vect{R}_{jli} \vect{Q}_{ji}^{-1} \bigg( \sum_{l'=1}^{L} \vect{h}_{jl'i} + \frac{1}{\sqrt{\rho^{\rm{tr}}}} \vect{n}_{ji}   \bigg) \!\sim \!\CN \left( \vect{0},  \vect{\Phi}_{jlli} \right)
\end{align}
where $\vect{n}_{ji} \sim \CN (\vect{0}, \vect{I}_{M})$, $\vect{Q}_{ji} = \sum_{l'=1}^{L} \vect{R}_{jl'i} + \frac{1}{\rho^{\rm{tr}}} \vect{I}_{M}$, and $\vect{\Phi}_{jlli}  = \vect{R}_{jli} \vect{Q}_{ji}^{-1} \vect{R}_{jli}$.
The estimation error $\tilde{\vect{h}}_{jli} = \vect{h}_{jli} - \hat{\vect{h}}_{jli}  \sim \CN \left( \vect{0}, \vect{R}_{jli}- \vect{\Phi}_{jlli} \right)$ is independent of $\hat{\vect{h}}_{jli}$. The mutual interference generated by the pilot-sharing UEs is known as \emph{pilot contamination} and has two main consequences in the channel estimation process \cite[Sec.~3.3.2]{massivemimobook}. The first is the reduced estimation quality, whereas the second is that the estimates $\hat{\vect{h}}_{j1i}, \ldots, \hat{\vect{h}}_{jLi}$ become correlated:
\begin{align}\label{theta_jmni}
\mathbb{E}\{ \hat{\vect{h}}_{jl'i} \hat{\vect{h}}_{jli}^{\Htran}\} =  \vect{\Phi}_{jl'li} = \vect{R}_{jl'i} \vect{Q}_{ji}^{-1} \vect{R}_{jli}. 
\end{align}
Both have an impact on the UEs' performance but it is only the second one that is responsible of the so-called \emph{coherent interference} \cite[Sec.~4.2]{massivemimobook}, which might increase linearly with $M$, just as the signal term. This is investigated later in detail.

We call ${\bf v}_{jk} \in \mathbb {C}^{M}$ the receive combining vector associated with UE $k$ in cell $j$. 
The uplink ergodic capacity can be lower bounded by the achievable spectral efficiency (SE)\cite{Marzetta2016a,massivemimobook}
\begin{equation} \label{eq:uplink-rate-expression-general}
\mathsf{SE}_{jk}^{\rm {ul}} = \left( 1- \frac{K}{\tau_c} \right) \mathbb{E} \left\{ \log_2  \left( 1 + \gamma_{jk} \right) \right\} \quad \textrm{[bit/s/Hz] }
\end{equation}
with the instantaneous effective SINR
\begin{align} \notag
\gamma_{jk}
& =  \frac{ |  \vect{v}_{jk}^{\Htran} \hat{\vect{h}}_{jjk} |^2  }{\!\!\!\!{\mathbb{E}}\left\{ 
\!\sum\limits_{(l,i)\ne (j,k)} \!\!\!\!| \vect{v}_{jk}^{\Htran} {\vect{h}}_{jli} |^2
+| \vect{v}_{jk}^{\Htran} \tilde{\vect{h}}_{jjk} |^2+  \frac{1}{\rho^{\rm{ul}}}\vect{v}_{jk}^{\Htran} \vect{v}_{jk}  
\Big| \{ \hat{\vect{h}}_{jli} \} \!\right\}} \\&
= \frac{ |  \vect{v}_{jk}^{\Htran} \hat{\vect{h}}_{jjk} |^2  }{ 
 \vect{v}_{jk}^{\Htran}  \left(   \sum\limits_{(l,i)\ne (j,k)}  \hat{\vect{h}}_{jli} \hat{\vect{h}}_{jli}^{\Htran} +   \vect{Z}_j + \frac{1}{\rho^{\rm{ul}}}  \vect{I}_{M}\right) \vect{v}_{jk}  
}   \label{eq:uplink-instant-SINR}
\end{align}
where ${\mathbb{E}}\{\cdot|\{ \hat{\vect{h}}_{jli} \} \}$ denotes the conditional expectation given the {MMSE} estimates $\{ \hat{\vect{h}}_{jli}:\forall l,i \} $ available at BS $j$ and 
\begin{align}
\vect{Z}_j = \sum_{l=1}^{L} \sum_{i=1}^{K} (\vect{R}_{jli} - \vect{\Phi}_{jlli}).
\end{align} 

\subsection{Optimal Receive Combining: M-MMSE}
For notational convenience, we define $\hat{\vect{H}}_{jk}\in\mathbb{C}^{M \times L}$ as
\begin{equation}
\hat{\vect{H}}_{jk} = [\hat{\vect{h}}_{j1k}, \hat{\vect{h}}_{j2k}, \ldots, \hat{\vect{h}}_{jLk}]
\end{equation}
the matrix collecting channel estimates of pilot sharing UEs and call $\hat{\vect{H}}_{jk}^{[j]}\in\mathbb{C}^{M\times (L-1)}$ the matrix obtained from $\hat{\vect{H}}_{jk} $ after removing the vector $\hat{\vect{h}}_{jjk}$. 

As shown in \cite{BHS18A,EmilEURASIP17}, the instantaneous effective {SINR} in \eqref{eq:uplink-instant-SINR} is a generalized Rayleigh quotient with respect to $\vect{v}_{jk}$ and thus is maximized by the M-MMSE combining vector: 
\begin{equation} \label{eq:MMSE-combining}
\vect{v}_{jk} =  \Bigg(  \sum\limits_{l=1}^L\sum\limits_{i=1}^K \hat{\vect{h}}_{jli} \hat{\vect{h}}_{jli}^{\Htran} + \vect{Z}_j+ \frac{1}{\rho^{\rm{ul}}}  \vect{I}_{M} \Bigg)^{\!-1}  \!\!  \hat{\vect{h}}_{jjk}.
\end{equation}
Plugging \eqref{eq:MMSE-combining} into \eqref{eq:uplink-instant-SINR} yields
\begin{align} \label{eq:gammajk_MMSE}
\!\!\!\!\!\!\gamma_{jk} &=  \hat{\vect{h}}_{jjk}^{\Htran} {\vect{U}}_{jk}^{-1} \hat{\vect{h}}_{jjk} \!\!\end{align}
where\vspace{-0.5cm}
\begin{align} \label{U_jk}
{\vect{U}}_{jk} = \hat{\vect{H}}_{jk}^{[j]}{(\hat{\vect{H}}_{jk}^{[j]})}^{\Htran}+ \overbrace{\sum_{l} \sum_{i\ne k} \hat{\vect{h}}_{jli}\hat{\vect{h}}_{jli}^{\Htran} \!+ \vect{Z}_j+ \frac{1}{\rho^{\rm{ul}}}  \vect{I}_{M}}^{\triangleq {\bf A}_{jk}}.
\end{align}
It can be shown that \eqref{eq:MMSE-combining} also minimizes ${\rm{MSE}}_{jk}^{\rm {ul}} = \mathbb{E} \{|s_{jk} - \vect{v}_{jk}^{\Htran}\vect{y}_j|^2\,|\, \{ \hat{\vect{h}}_{jli} \}\}$
which represents the conditional MSE between the data signal $s_{jk}$ and the received signal $\vect{v}_{jk}^{\Htran}\vect{y}_j$ after receive combining. By using standard calculus, \eqref{eq:gammajk_MMSE} can be equivalently expressed as
\begin{align} \label{eq:gammajk_MMSE_1}
\!\!\gamma_{jk} =  \frac{1}{{\rm{MSE}}_{jk}^{\rm {ul}}}  -1\end{align}
where ${\rm{MSE}}_{jk}^{\rm {ul}}$ (as obtained after plugging \eqref{eq:MMSE-combining} into its definition) reads
\begin{align} {\rm{MSE}}_{jk}^{\rm {ul}} = \left[\left(  {\bf I}_{L}+ \hat{\vect{H}}_{jk}^{\Htran}{\bf A}_{jk}^{-1}\hat{\vect{H}}_{jk}\right)^{-1}\right]_{j,j}.\label{eq:appendixc_7.1}
\end{align}
Notice that the right-hand-side of \eqref{eq:appendixc_7.1} can be rewritten in many equivalent forms by collecting the channel estimate vectors in \eqref{U_jk} in different matrices. The reason that we consider the form in \eqref{eq:appendixc_7.1} is that ${\bf A}_{jk}$ is
independent of $\hat{\vect{H}}_{jk}$. This not only makes the asymptotic analysis of \eqref{eq:gammajk_MMSE_1} rather simple (as shown later) but also allows to gain the following interesting insights. By using the same steps as in \cite[Eq. (8)]{Paul2006}, \eqref{eq:gammajk_MMSE_1} can be equivalently rewritten as in \eqref{eq:appendixc_7} at the top of the page, which is obtained as the difference between two terms. The first depends on the inverse of the matrix ${\bf A}_{jk}$ defined in \eqref{U_jk}, which is obtained from all the UE channels that do \emph{not} cause pilot contamination to UE $k$ in cell $j$. The second term in \eqref{eq:appendixc_7} depends not only on ${\bf A}_{jk}$ but also on the channel estimates of all the pilot-sharing UEs, which enters into $\hat{\vect{H}}_{jk}^{[j]}$. Therefore, it can be seen as the loss induced in the effective instantaneous SINR by the correlation among pilot contaminating channels. Notice that, although independent from $\hat{\vect{H}}_{jk}^{[j]}$, the first term is also affected by pilot contamination due to the reduced channel estimation quality. As shown later by simulations, both terms grow with $M/K$ when $M,K\to \infty$.


Table~\ref{tab:cost_linear_processing} summarizes the total complexity for evaluating \eqref{eq:gammajk_MMSE} and \eqref{eq:gammajk_MMSE_1} (in terms of number of complex multiplications) for each coherence block, under the assumption that the statistical matrices $\{{\bf Z}_j\}$ and $\{\vect{R}_{jli}, {\bf{Q}}_{ji}^{-1}\}$ are precomputed and stored at the BSs. Clearly, the computation of the effective SINR is very involved in all cases. In particular, the complexity scales as $M^3$ and $M^2K$, which are basically the same when $M$ and $K$ grow with a fixed ratio. Notice also that these operations must be performed over hundreds of coherence blocks to obtain a good estimate of the SE as given by \eqref{eq:uplink-rate-expression-general}. This makes it hard to evaluate the SE when $M$ and $K$ grow large, as envisioned in future Massive MIMO networks. Nevertheless, the evaluation of the effective SINR can be crucial for both physical layer (link-level) and network layer (system-level) simulations and optimization. While the former aims at investigating issues such as adaptive modulation and coding, feedback, channel encoding and decoding, the latter focuses on network-related issues such as scheduling and mobility management \cite{Ikuno2010}.

\begin{table}[t]  \vspace{0.2cm}      \caption{Number of complex multiplications per coherence block to compute \eqref{eq:gammajk_MMSE} and \eqref{eq:gammajk_MMSE_1}.\vspace{-.1cm}}  \label{tab:cost_linear_processing}
\centering
    \begin{tabular}{|c|c|c|} 
    \hline
    {} & {\!\!\!Channel estimation\!\!\!}&{\!\!\!\!\!Computation of $\gamma_{jk}$\!\!\!\!\!}   \\      \hline\hline 
    \!\!\eqref{eq:gammajk_MMSE}\!\! &   $M\tau_p + LM^2 $ & $\frac{M^2 + M}{2}(LK+1)+\frac{M^3 - M}{3}$   \\   \hline  
    \!\!\eqref{eq:gammajk_MMSE_1}\!\! &   $M\tau_p + LM^2 $ & $\frac{M^2 + M}{2}(L^2(K+2)+L)+\frac{M^3 - M}{3}+\frac{L^3 - L}{3}$ \!\!\!\!\!\!   
%
%
  
   \\ \hline
  
    \end{tabular}\vspace{-0.5cm}
\end{table}
\section{Asymptotic Analysis}
As mentioned in the introduction, we want to analyze $\gamma_{jk}$ in the regime where $M,K\to \infty$ with $\liminf M/K>0$, which might provide better approximations of practical setups where both $M$ and $K$ are large. To this end, we assume that $\rho^{\rm{ul}} = \rho/M$ with $\rho$ being fixed and make the following assumptions.
\begin{assumption}\label{assumption_3}As $M\to \infty$ $\forall j,l,i$, $\liminf_M \;\frac{1}{{M}}\tr ( \vect{R}_{jli}) > 0 $ and $
	\limsup_M \;\| \vect{R}_{jli}\|_2 < \infty$.
\end{assumption}
These conditions are widely used for the asymptotic analysis \cite{Hoydis2013,massivemimobook} of Massive MIMO. The first implies that the array gathers more energy as $M$ increases, whereas the second implies that the energy is spread over many spatial dimensions. 
 \setcounter{equation}{13}
For convenience, we define 
 \begin{align}\label{T_j}
 {\bf T}_{j}^{\star} = \left(\frac{1}{M}\sum\limits_{l=1}^{L}\sum\limits_{i=1}^{K}\frac{\vect{\Phi}_{jlli} }{1+ \mu_{jli}^{\star}} + \frac{1}{M}\vect{Z}_j + \frac{1}{\rho}  \vect{I}_{M}\right)^{-1}
  \end{align}
  where the coefficients $\{\mu_{jli}^{\star} : \forall l,i\}$ are solutions of the following system of equations:
    \begin{align}\label{mu_jlk}
\!\!\!\! \mu_{jlk} \!= \!\frac{1}{M}\tr  \!\left( \!\vect{\Phi}_{jllk} \!\left(\frac{1}{M}\sum\limits_{l=1}^{L}\sum\limits_{i=1}^{K}\frac{\vect{\Phi}_{jlli} }{1+\mu_{jli}} + \frac{1}{M}\vect{Z}_j+ \frac{1}{\rho}  \vect{I}_{M}\!\!\right)^{\!\!\!-1}\!\right).\!\!
  \end{align}
Moreover, we define $\vect{B}_{jk}\in \mathbb{C}^{L \times L}$ with entries
 \begin{align}\label{sigma_jk_1}
\big[\vect{B}_{jk} \big]_{l,l'}=\frac{1}{M}\tr \left(  \vect{\Phi}_{jl'lk} {\bf T}_{j}^{\star}\right)\end{align}
 where $ \vect{\Phi}_{jl'lk} $ is given by \eqref{theta_jmni}, and denote by $\vect{B}_{jk}^{[jj]}\in \mathbb{C}^{(L-1) \times (L-1)}$ the matrix obtained from $\vect{B}_{jk}$ after removing the $j$th column and $j$th row. Also, $\vect{b}_{jk}^{[j]}\in \mathbb{C}^{L-1}$  is obtained from the $j$th column of $\vect{B}_{jk}$ after removing $[\vect{B}_{jk}]_{j,j}$. 

\begin{theorem} \label{theorem:M-MMSE}
If Assumptions~\ref{assumption_3} holds and M-MMSE combining is used with $\rho^{\rm{ul}} = \rho/M$, then 
\begin{align}\label{eq:general result}
\!\!\!\gamma_{jk}\asymp \overline{\gamma}_{jk} &= \frac{1}{\left[\left(  {\bf I}_{L}+ \vect{B}_{jk}\right)^{-1}\right]_{j,j}}-1\\&= [\vect{B}_{jk}]_{j,j} - \underbrace{{\big(\vect{b}_{jk}^{[j]}\big)}^{\Htran}\left(  {\bf I}_{L-1}+ \vect{B}_{jk}^{[jj]}\right)^{-1}\!\!\!\vect{b}_{jk}^{[j]}}_{\triangleq  \overline{\zeta}_{jk}}\label{eq:general result_1}
\end{align}
when $M,K\to \infty$ with $\liminf M/K>0$. \end{theorem}
\begin{IEEEproof}
The proof of \eqref{eq:general result} is given in the appendix by applying tools from random matrix theory to \eqref{eq:gammajk_MMSE_1}. Simple arguments (e.g., \cite{Paul2006}) can be used to obtain \eqref{eq:general result_1} from \eqref{eq:general result}, which can be seen as an asymptotic approximation of \eqref{eq:appendixc_7}. Interestingly, the asymptotic analysis is much simpler than that for S-MMSE \cite{Hoydis2013}, where similar tools can be used. This is because with S-MMSE, $\gamma_{jk}$ does not reduce to the quadratic form in \eqref{eq:gammajk_MMSE} (from which \eqref{eq:gammajk_MMSE_1} follows) as with M-MMSE, and thus an asymptotic approximation can only be obtained by deriving asymptotic expressions for each single term in \eqref{eq:uplink-instant-SINR}. This latter approach was also taken in \cite{EmilEURASIP17}, even though M-MMSE was considered (but for uncorrelated channels).
\end{IEEEproof}
\smallskip
Theorem \ref{theorem:M-MMSE} provides asymptotic approximations of $\gamma_{jk}$ that are deterministic and thus can be inserted into \eqref{eq:uplink-rate-expression-general} to directly obtain approximations of the SE, without the need to evaluate the expectation.
The computation requires first to obtain the coefficients $\{\mu_{jli}^{\star}: \forall j\}$ by solving $L$ sets of $KL$ fixed-point equations. In \cite{Wagner2012}, it is proved that $\{\mu_{jli}^{\star}:\forall j\}$ can be efficiently obtained by an iterative algorithm, which needs only a few iterations to converge. We notice that $\{\mu_{jli}^{\star}\}$ only depend on the channel statistics and, therefore, can be precomputed and only updated when the channel statistics change substantially (e.g., due
to UE mobility or new scheduling decisions).

%

\begin{table*}[t]
\renewcommand{\arraystretch}{1.}
\centering
\caption{Network parameters}
\label{table:system_parameters_running_example}
\begin{tabular}{|c|c|}
\hline

      Cell area (with wrap around) &  $0.4$\,km $ \times\,  0.4$\,km \\
  Number of cells&  $L = 4$ \\

%

Samples per coherence block & $\tau_c = 200$ \\


Distance between UE $k$ in cell $l$ and BS~$j$ & $d_{lk}^{\,j}$ \\

\begin{tabular}{@{}c@{}} Large-scale fading coefficient for \\ the channel between UE $k$ in cell $l$ and BS~$j$\end{tabular}
& $\beta_{lk}^{j} =  -148.1 - 37.6 \, \log_{10} \left( \frac{d_{lk}^{j}}{1\,\textrm{km}} \right) + F_{lk}^{j}$\,dB\\
 
Shadow fading between UE $k$ in cell $l$ and BS~$j$ & $F_{lk}^{j} \sim \mathcal{N}(0,10)$ \\




\hline
\end{tabular}\vspace{-0.5cm}
\end{table*}

Once $\{\mu_{jli}^{\star}\}$ are computed, we need roughly $\frac{4M^3 - M}{3}KL^2$ complex multiplications to compute \eqref{eq:general result}, which is not too different from the complexity of computing \eqref{eq:gammajk_MMSE} and \eqref{eq:gammajk_MMSE_1} (see Table~\ref{tab:cost_linear_processing}). The key difference is that the latter ones need to be computed for every channel realization (or at least very many realizations to approximate the expectation in \eqref{eq:uplink-rate-expression-general} by Monte Carlo simulations). Hence, the asymptotic approximation $\overline{\gamma}_{jk}$ will substantially reduce the computational burden. Moreover, the numerical results in Section IV prove that it is both asymptotically tight and accurate for systems with finite dimensions. 

In the appendix, it is shown that the two terms in \eqref{eq:general result_1} can be bounded as follows:
\begin{align}\label{A.3}
 &\!\!\!\!\!\!\frac{M}{KL} \frac{1}{\varsigma} \le \frac{\big[\vect{B}_{jk} \big]_{j,j}}{\frac{1}{M}\tr \left(  \vect{\Phi}_{jjjk} \right)}\le \frac{M}{KL}\frac{1}{\eta}
\end{align} 
and
\begin{align}\label{eq:bound}
\!\! \frac{(\frac{M}{KL})^2 \varsigma^\prime}{1 + \frac{M}{KL} \eta^\prime} \le \overline{\zeta}_{jk}\le \frac{(\frac{M}{KL})^2 \varsigma^\prime}{1 + \frac{M}{KL}  \frac{1}{L-1}\eta^\prime}
\end{align}
where $\eta,\eta^\prime,\varsigma$ and $\varsigma^\prime$ are defined in the appendix. As seen, both terms are bounded below and above by $M/K$ (up to constant factors), as validated later by numerical results.



\begin{remark}[Orthogonal correlation matrices]
It is known that the SE increases when the interfering UEs' have different spatial correlation properties \cite{massivemimobook}. This is confirmed by the expression in \eqref{eq:general result}. In the extreme case of $\vect{R}_{jl'k} \vect{R}_{jlk} = {\bf 0}_{M}$ $\forall l'\ne l$,
we have that 
$\vect{B}_{jk}$ becomes diagonal and thus  
\begin{align}
\gamma_{jk}\asymp \frac{1}{M}\tr \left(  \vect{\Phi}_{jjjk} {\bf T}_{j}^{\star}\right) = \mu_{jjk}^{\star}
\end{align}
where $\mu_{jjk}^{\star}$ is obtained from \eqref{mu_jlk} after replacing $\vect{\Phi}_{jlli}$ with $\vect{\Phi}_{jlli} = \vect{R}_{jli} \big(\vect{R}_{jli} + \frac{1}{\rho^{\rm{tr}}} \vect{I}_{M}\big)^{-1} \vect{R}_{jli}$\cite[Lemma B.6]{massivemimobook},
which does not depend on the pilot-sharing UEs. A similar result holds if $\{\vect{R}_{jl'k}: \forall l'\ne l\}$ are asymptotically spatially orthogonal $\frac{1}{M}\tr \big(\vect{R}_{jl'k} \vect{R}_{jlk}\big) \asymp 0$. This implies that the loss due to correlation among pilot contaminating channels in \eqref{eq:appendixc_7} can be avoided if their correlation matrices are (asymptotically) spatially orthogonal. However, this condition only appears in special cases \cite{massivemimobook} and thus the SINR will always be affected by pilot contamination in practice.
\end{remark}

\section{Numerical results}
The asymptotic analysis is now validated by using the network setup in Table~\ref{table:system_parameters_running_example}. 
Each BS is equipped with a uniform linear array with half-wavelength antenna spacing. The correlation matrices $\{\vect{R}_{li}^j\}$ are generated by using the exponential correlation model with correlation factor $r =0.5$ between adjacent antennas. The large-scale fading coefficient $\beta_{li}^j$ is reported in Table~\ref{table:system_parameters_running_example}.  The normalized transmit power is $\rho=114$ dBm, while $\rho^{\rm{tr}}=\rho K$.

\begin{figure}[t!]
\begin{center}
\begin{overpic}[unit=1mm,width=1.1\columnwidth]{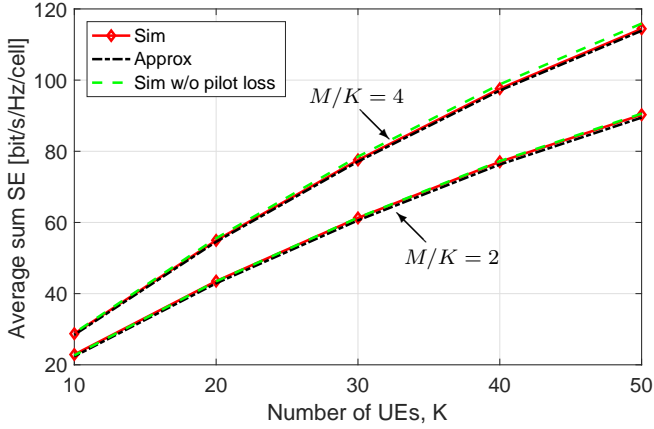}
\put(58,22){{\footnotesize$M/K=2$}}
\put(61,25){\vector(-1,1){4}}
\put(45,44){{\footnotesize $M/K=4$}}
\put(52,43){\vector(1,-1){4}}
\end{overpic}
\end{center} \vspace{-0.5cm}
\caption{Average UL sum SE with M-MMSE combining as a function of $K$, when $M$ increases with $K$ with  fixed antenna-UE ratios $M/K$.} \vspace{-0.3cm}\label{figure:ul_se_vs_MK}
\end{figure}

\begin{figure}[t!]
\begin{center}
\begin{overpic}[unit=1mm,width=1.1\columnwidth]{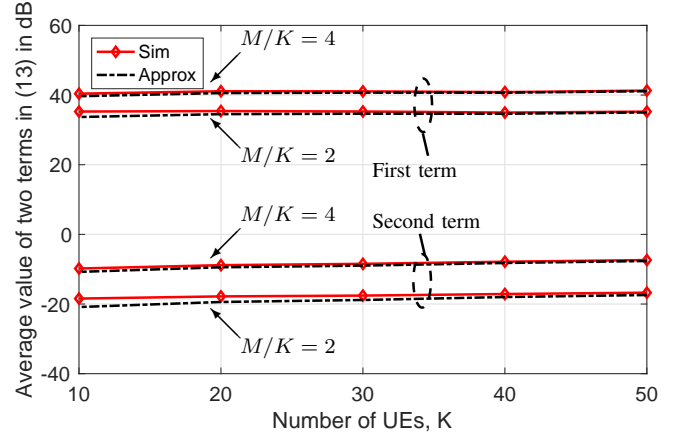}
\put(35,11){{\footnotesize$M/K=2$}}
\put(35,13){\vector(-1,1){4}}
\put(35,29){{\footnotesize $M/K=4$}}
\put(35,28){\vector(-1,-1){4}}
\put(35,37){{\footnotesize$M/K=2$}}
\put(35,39){\vector(-1,1){4}}
\put(35,53){{\footnotesize $M/K=4$}}
\put(35,52){\vector(-1,-1){4}}
\put(53,35){{\footnotesize First term}}
\put(53,28){{\footnotesize Second term}}
\end{overpic}
\end{center}\vspace{-0.5cm}
\caption{Average strength of the two terms in \eqref{eq:appendixc_7} in dB as a function of $K$ with fixed antenna-UE ratios $M/K$.} \vspace{-0.3cm}\label{figure:ul_se_vs_MK}
\end{figure}

\begin{figure*}[h] \setcounter{equation}{24}
  \begin{align}\label{A.boundT}
\frac{M}{KL}\frac{1}{\underbrace{\max_{jli}||\vect{\Phi}_{jlli}||_2 + \max_{jli}||\vect{R}_{jli} - \vect{\Phi}_{jlli}||_2 + \frac{1}{KL}\frac{1}{\rho}}_{\triangleq\varsigma}}{\bf I}_M \preceq {\bf T}_j^{\star} \preceq \frac{M}{KL}\frac{1}{\underbrace{\min_{jli} \lambda_{\min}(\vect{R}_{jli} - \vect{\Phi}_{jlli})+ \frac{1}{KL}\frac{1}{\rho}}_{\triangleq\eta}}{\bf I}_M
\end{align}\vspace{-0.2cm}
\hrule\vspace{-0.5cm}
  \end{figure*}

Fig. \ref{figure:ul_se_vs_MK} plots the average sum SE per cell as a function of $K$ when $M$ is increased proportionally to $K$ with $M/K=2,4$. The curve `Sim' refers to the SE obtained with M-MMSE by Monte Carlo simulations, while 'Approx' is computed by means of the asymptotic approximation provided in Theorem \ref{theorem:M-MMSE}. As seen, the SE obtained with the asymptotic approximation perfectly matches the Monte Carlo simulations in all investigated cases. While the SINR (not shown for space limitations) grows linearly with $K$ in both cases, the SE starts decreasing because of the pilot overhead, which enters in \eqref{eq:uplink-rate-expression-general} through the pre-log factor. To quantify the impact of the SINR loss caused by the pilot-contaminating UEs, we also report the SE as obtained with \eqref{eq:appendixc_7} after neglecting the second term.  Only a negligible difference is observed. This means that the correlation among channel estimates of pilot-sharing UEs has a very minor impact on SE. 

To validate the scaling behaviour of the two terms in \eqref{eq:appendixc_7} and quantify their relative importance, Fig. \ref{figure:ul_se_vs_MK} plots their average values in dB for an arbitrary UE in the cell. The results obtained with the asymptotic approximations in \eqref{eq:general result_1} perfectly match the Monte Carlo simulations. Moreover, both maintain constant as $K$ grows but increases with $M/K$. The first term is roughly $40-50$ dB higher than the second one for both antenna-UE ratios. Although the situation is different if a specific UE in the cell is considered, the loss caused by the correlation among pilot contaminating channels is always several dBs lower. This implies that it has a minor impact compared to intra- and inter-cell interference.

\section{Conclusions}

We analyzed Massive MIMO in the traditional large-system limit where the number of antennas and UEs are growing with a fixed ratio, which is different from the ``Marzetta limit'' where only the number of antennas grows. We provided an asymptotically tight low-complexity approximation of the uplink SINR in Massive MIMO networks with the optimal M-MMSE combiner and arbitrary correlated Rayleigh fading channels. Numerical results were used to validate the high accuracy of this approximation for realistic system dimensions. When applied to practical networks, such a result can be used to evaluate the SE of network and/or the effective SINR without to carry out extensive Monte Carlo simulations.
In particular, expressions like this are valuable for resource allocation and optimization, as exemplified in \cite{EmilEURASIP17}.

\section*{Appendix}

\setcounter{equation}{20}
Since ${\bf A}_{jk}$ is
independent of $\hat{\vect{H}}_{jk}$ and $\hat{\vect{h}}_{jlk}\sim  \CN \left( \vect{0},  \vect{\Phi}_{jllk} \right)$, under Assumption~\ref{assumption_3} from the trace lemma \cite{Wagner2012} it follows that\footnote{Note that it can be shown that the matrices $\vect{\Phi}_{jllk} $ have uniformly bounded spectral norm due to Assumption \ref{assumption_3}.}
\begin{align}\notag
\left[\frac{1}{M}\hat{\vect{H}}_{jk}^{\Htran}\tilde{\bf A}_{jk}^{-1}\hat{\vect{H}}_{jk}\right]_{l,l'}&=\frac{1}{M}\hat{\vect{h}}_{jlk}^{\Htran}\tilde{{\bf A}}_{j,\setminus k}^{-1}\hat{\vect{h}}_{jl'k}  \\&\asymp \frac{1}{M}\tr  \bigg(  \vect{\Phi}_{jl'lk} \tilde{\bf A}_{jk}^{-1} \bigg)
 \end{align}
 with $\tilde{\bf A}_{jk} =  \frac{1}{M}{\bf A}_{jk}$ and $ \vect{\Phi}_{jl'lk} $ given by \eqref{theta_jmni}.
By using \cite[Th.~1]{Hoydis2013} under Assumption~\ref{assumption_3}, we obtain 
\begin{align}
\frac{1}{M}\tr  \left(   \vect{\Phi}_{jl'lk}\tilde{\bf A}_{jk}^{-1} \right)\asymp\big[\vect{B}_{jk} \big]_{l,l'}
 \end{align}
where the entries of $\vect{B}_{jk}$ are defined in \eqref{sigma_jk_1}.
Since each of the entries of $\frac{1}{M}\hat{\vect{H}}_{jk}^{\Htran}\tilde{\bf A}_{jk}^{-1}\hat{\vect{H}}_{jk}$ converges, we have that
  \begin{align}
\left\|\frac{1}{M}\hat{\vect{H}}_{jk}^{\Htran}\tilde{\bf A}_{jk}^{-1}\hat{\vect{H}}_{jk} - \vect{B}_{jk}  \right\|_2\asymp 0
  \end{align}
from which it follows that
  \begin{align}
\left\|\Big({\bf{I}}_L + \frac{1}{M}\hat{\vect{H}}_{jk}^{\Htran}\tilde{\bf A}_{jk}^{-1}\hat{\vect{H}}_{jk}\Big)^{-1} - \Big( {\bf{I}}_L + \vect{B}_{jk} \Big)^{-1} \right\|_2\asymp 0.
  \end{align}
Plugging this result into \eqref{eq:gammajk_MMSE_1} we obtain \eqref{eq:general result} from the continuous mapping theorem.
  
  Under Assumption~\ref{assumption_3}, the matrix ${\bf T}_j^{\star}$ can be bounded as in \eqref{A.boundT} at top of the page. Hence, from \eqref{sigma_jk_1} we have that 
\setcounter{equation}{25}
\begin{align}\label{A.3}
 &\!\!\!\!\!\!\frac{M}{KL} \frac{1}{\varsigma}\frac{1}{M}\tr \left(  \vect{\Phi}_{jllk} \right) \le \big[\vect{B}_{jk} \big]_{l,l}\le \frac{M}{KL}\frac{1}{\eta}\frac{1}{M}\tr \left(  \vect{\Phi}_{jllk} \right).
\end{align}
For the second term in \eqref{eq:general result_1}, we notice that
\begin{align}\label{A.4}
 \frac{1}{L-1} \tr \left(  \vect{B}_{jk}^{[jj]} \right) \vect{I}_{L-1}\preceq \vect{B}_{jk}^{[jj]} \preceq \tr \left(  \vect{B}_{jk}^{[jj]} \right) \vect{I}_{L-1}.
\end{align}
By using \eqref{A.3} and \eqref{A.4} with ${\bf x}^{\Htran}\vect{A}{\bf x}^{\Htran} \le {\bf x}^{\Htran}\vect{C}{\bf x}^{\Htran}$ if $\vect{C}-\vect{A} \succeq \vect{0} $, we thus obtain \eqref{eq:bound} with $\eta^\prime= \frac{1}{\eta} \sum\nolimits_{l=1,l\ne j}^{L} \frac{1}{M}\tr\left(  \vect{\Phi}_{jllk} \right)$ and $\varsigma^\prime= \frac{1}{\varsigma^2} \sum\nolimits_{l=1,l\ne j}^{L} \big(\frac{1}{M}\tr (  \vect{\Phi}_{jllk} )\big)^2.$
    \bibliographystyle{IEEEtran}
\bibliography{IEEEabrv,ref}
\end{document}